\documentclass[
showpacs,preprintnumbers,amsmath,amssymb]{revtex4}
\usepackage{amsmath}
\usepackage{graphicx}
\begin{document}

\tolerance=5000

\def\be{\begin{equation}}
\def\ee{\end{equation}}
\def\bea{\begin{eqnarray}}
\def\eea{\end{eqnarray}}
\def\tr{{\rm tr}\, }
\def\nn{\nonumber \\}
\def\e{{\rm e}}
\title{Modified f(R) gravity from scalar-tensor theory and inhomogeneous EoS dark energy }

\author{Diego S\'{a}ez-G\'{o}mez$^{1,}$\footnote{Electronic address:saez@ieec.uab.es}}
\affiliation{$^{1}$Institut de Ci\`{e}ncies de l'Espai
ICE/CSIC-IEEC, Campus UAB, Facultat de Ci\`encies, Torre
C5-Parell-2a pl, E-08193 Bellaterra (Barcelona) Spain}
\begin{abstract}
The reconstruction of f(R)-gravity is showed by using an auxiliary scalar field in the context of cosmological evolution, this development provide a way of reconstruct the form of the function $f(R)$ for a given evolution of the Hubble parameter. In analogy, f(R)-gravity may be expressed by a perfect fluid with an inhomogeneous equation of state (EoS) that depends on the Hubble parameter and its derivatives. This mathematical equivalence that may confuse about the origin of the mechanism that produces the current acceleration, and possibly  the whole evolution of the Hubble parameter, is shown here.   
\end{abstract}
\maketitle
\section{Introduction}
Ever since the accelerated nature of the dynamics of the Universe was discovered in 1998 by the observations of SN Ia, a lot of theoretical descriptions have been proposed such that the observational data is satisfied. They have established that the EoS parameter $w$ for the fluid that governs the Universe is close to -1.  The majority of these models are included in the so-called dark energy models, where the origin of the dark energy is a scalar field or a perfect fluid with an inhomogeneous equation of state(EoS),  which should be the responsable for the current acceleration, and may be even on the whole expansion history of the Universe. On the other hand, the modified f(R) theories of gravity avoid the need to introduce dark energy, and may give an explanation about the origin of the current accelerated expansion and even on the expansion history of the Universe(for recent reviews, see \cite{f(R)review1} and \cite{f(R)review2}). \\
In this sense, the cosmic acceleration and the cosmological properties of metric formulation f(R) theories have been studied in Refs.\cite{f(R)1}-\cite{f(R)54}. Recently the main focus has been improviding a f(R)-theory that reproduces the whole history of the Universe, including the early accelerated epoch (inflation) and  the late-time acceleration at the current epoch (see \cite{F(R)toScalar}-\cite{F(R)UnfInfCosAcceAndSingularity}), where the possible future singularities have been studied in the context of f(R)-gravity(see \cite{F(R)UnfInfCosAcceAndSingularity}). It is important to remark that the main problem that this kind of theories found at the begining of its development was the local gravitational test; nowadays several viable models have been proposed, which pass the solar system tests and reproduce the cosmological history (see \cite{F(R)UnfInfCosAcce2andsolartest}-\cite{f(R)viable10}). \\
In the present paper, the reconstruction of f(R)-gravity is shown, to be possible in the cosmological context by using an auxiliary scalar field and then  various examples are given where the current accelerated expansion is reproduced and also  the whole history of the Universe. This kind of reconstruction is well done by using an scalar field without kinetic term, which differs of the kind of quintessence reconstruction models (see \cite{Quintessence1} and \cite{Quintessence2}) where the scalar field presented has a non-zero kinetic term. Also it is investigated the analogy between the so-called dark fluids, whose EoS is inhomogeneous and which have been proposed as candidate of dark energy (see \cite{InhEoS1}-\cite{MyW2}), and the f(R)-theories, a reconstruction of such kind of theories is shown by using such type of perfect fluids.\\
The paper is organized in two sections, in the first one the reconstruction of f(R)-gravity is obtained by using scalar-tensor theory. In the second section, the analogy between the considerations of dark fluids with an EoS depending on the Hubble parameter and its derivatives, and  f(R)-gravity is considered.  
\section{Reconstruction of f(R)-gravity}

In this section, it will be shown how f(R) theory may be reconstructed in such a way that cosmological solutions can be  obtained. Let us start with the action for f(R)-gravity:
\be
S=\int d^4x\sqrt{-g}\left(f(R)+L_m\right)\ . \label{1.1}
\ee
Here $L_m$ denotes the lagrangian of some kind of matter. The field equations are obtained by varying the action on $g_{\mu\nu}$, then they are  given by:
\be
R_{\mu\nu}f'(R)-\frac{1}{2}g_{\mu\nu}f(R)+g_{\mu\nu}\Box f'(R)-\nabla_{\mu}\nabla_{\nu}f'(R)=\frac{\kappa^2}{2}T^{(m)}_{\mu\nu}\ ,
\label{1.2}
\ee
where $T^{(m)}_{\mu\nu}$ is the energy-momentum tensor for the matter that filled the Universe. We assume a flat FRW metric:
\be
ds^2=-dt^2+a^2(t)\sum^{3}_{i=1}dx_{i}^2 \ .
\label{1.2a}
\ee
Then, if $T^{(m)}_{\mu\nu}$ is a perfect fluid, the modified Friedmann equations  for the Hubble parameter $H(t)=\frac{\dot{a}}{a}$, take the form:
\bea
\frac{1}{2}f(R)-3(H^2+\dot{H})f'(R)+18f''(R)(H^2\dot{H}+H\ddot{H})=\frac{\kappa^2}{2}\rho_{m}\ , \nn
\frac{1}{2}f(R)-(3H^2+\dot{H})f'(R)-\Box f'(R)=-\frac{\kappa^2}{2}p_m\ ,   
\label{1.3}
\eea
where the Ricci scalar is given by $R=6(2H^2+\dot{H})$. Hence,  by the equations (\ref{1.3}), any cosmology may be reproduced for a given function $f(R)$. Nevertheless, in general it is very difficult to get an exact cosmological solution directly from (\ref{1.3}). It is a very useful  technique developed in \cite{f(R)4} and \cite{F(R)toScalar}, where an auxiliary scalar field without kinetic term is introduced, then the action (\ref{1.1}) is rewritten as follows:
\be
S=\int d^4x\sqrt{-g}\left(P(\phi)R+Q(\phi)\right)\ ,
\label{1.4}
\ee
where the scalar field $\phi$ has no kinetic term. By  variation on the metric tensor $g_{\mu\nu}$ the field equation is obtained:
\be
\frac{1}{2}g_{\mu\nu}\left( P(\phi)R+Q(\phi)\right)+P(\phi)R_{\mu\nu}+g_{\mu\nu}\Box P(\phi)-\nabla_{\mu}\nabla_{\nu}P(\phi)= \frac{\kappa^2}{2}T^{(m)}_{\mu\nu}\ .
\label{1.5}
\ee
The action (\ref{1.4}) gives an additional equation for the scalar field $\phi$, obtained directly from the action by varying it with respect to $\phi$:
\be
P'(\phi)R+Q'(\phi)=0\ ,
\label{1.6}
\ee   
here the primes denote derivatives respect $\phi$. This equation may be resolved  with the scalar field as a function of $R$, $\phi=\phi(R)$, and then, replacing this result in the action (\ref{1.4}), the action (\ref{1.1}) is recovered,
\be
f(R)=P\left(\phi(R)\right)R+Q\left(\phi(R)\right)\ .
\label{1.7}
\ee
Hence, any cosmological model could be solved by the field equation (\ref{1.5}), and then by (\ref{1.6}) and (\ref{1.7}) the function $f(R)$ is obtained. For the metric (\ref{1.2a}), the Friedmann equations read:
\bea
3H\frac{dP(\phi)}{dt}+3H^2P(\phi)+\frac{1}{2}Q(\phi)-\frac{\rho_m}{2}=0\ , \nn
2\frac{d^2P(\phi)}{dt^2}+4H\frac{dP(\phi)}{dt}+(4\dot{H}+6H^2)P(\phi)+p=0\ .
\label{1.8}
\eea
We redefine the scalar field such that it is chosen to be the time coordinate $\phi=t$. The perfect fluid define by the energy-momentum tensor $T^{(m)}_{\mu\nu}$ may be seen as a sum of the different components (radiation, cold dark matter,..) which filled our Universe and whose  equation of state (EoS) is given by $p_m=w_m\rho_m$, then by the energy momentum conservation $\dot{\rho}_m +3H(1+w_m)\rho_m=0$, it gives:
\be
\rho_m=\rho_{m0}\exp\left(-3(1+w_m)\int dtH(t)\right)\ .
\label{1.9}
\ee
Hence, taking into account the equations (\ref{1.8}) and (\ref{1.9}) the Hubble parameter may be calculated as a function of the scalar field $\phi$, $H=g(\phi)$. By combining the equations (\ref{1.8}), the function $Q(\phi)$ is deleted, and it yields:
\be
2\frac{d^2P(\phi)}{d\phi^2}-2g(\phi)\frac{dP(\phi)}{d\phi}+4g'(\phi)P(\phi)+(1+w_m)\exp\left[-3(1+w_m)\int d\phi g(\phi) \right] =0\ .
\label{1.10}
\ee
By resolving this equation for a given function $P(\phi)$, a cosmological solution $H(t)$ is found, and the function $Q(\phi)$ is obtained by means of the equation given by (\ref{1.8}):
\be
Q(\phi)=-6(g(\phi))^2P(\phi)-6g(\phi)\frac{dP(\phi)}{dt}\ .
\label{1.11}
\ee
If we neglect the contribution of matter, then the equation (\ref{1.10}) is a first order differential equation on $g(\phi)$, and it can be easily resolved. The solution found is the following:
\be
g(\phi)=-\sqrt{P(\phi)}\int d\phi \frac{P''(\phi)}{2P^{2/3}(\phi)}+ kP(\phi)\ ,
\label{1.12}
\ee
where $k$ is an integration constant. As an example to show this construction, let us choose the following function that, as it is showed below, reproduce  late-time acceleration:
\be
P(\phi)=\phi^\alpha\ , \quad \mbox{where} \quad \alpha>1\ .
\label{1.13}
\ee
Then, by the result (\ref{1.12}), the following solution is found:
\be
g(\phi)=k\phi^{\alpha/2}+\frac{\alpha(\alpha-1)}{\alpha+2}\frac{1}{\phi}\ ,
\label{1.14}
\ee
where $k$ is an integration constant. By the expression (\ref{1.11}), the function $Q(\phi)$ is given by:
\be
Q(\phi)=-6\left[(k\phi)^2+\left(k+\frac{\alpha(2\alpha+1)}{\alpha+2} \right)\phi^{\frac{3\alpha-2}{2}}+\frac{\alpha^2(\alpha-1)(2\alpha+1)}{(\alpha+2)^2}\phi^{\alpha-2}  \right]\ .
\label{1.15}
\ee
The function (\ref{1.14}) gives the following expression for the Hubble parameter:
\be
H(t)=kt^{\alpha/2}+\frac{\alpha(\alpha-1)}{\alpha+2}\frac{1}{t}\ .
\label{1.16}
\ee
This solution may reproduce a Universe that passes through two phases for a conveniently choice of $\alpha$. For small times the Hubble and the  acceleration parameter take the expressions:
\bea
H(t)\sim\frac{\alpha(\alpha-1)}{\alpha+2}\frac{1}{t}\ , \nn
\frac{\ddot{a}}{a}\sim -\frac{\alpha(\alpha-1)}{\alpha+2}\left(1-\frac{\alpha(\alpha-1)}{\alpha+2} \right) \frac{1}{t^2}\ ,
\label{1.17}
\eea
where if $1+\sqrt{3}<\alpha\leq2$, the Universe is in a decelerated epoch for small times, which may be interpreted as the radiation/matter dominated epochs. When $t$ is large, the Hubble parameter takes the form:
\be
 H(t)=kt^{\alpha/2}\ .
\label{1.18}
\ee
This clearly gives an accelerated expansion that coincides with the current expansion that Universe experiences nowadays. Finnally, the expression for $f(R)$ (\ref{1.7}) is calculated by means of (\ref{1.13}) and (\ref{1.15}), and by the expression of the Ricci scalar $R=6(2g^2(\phi)+g(\phi))$, which is used to get $\phi(R)$. For simplicity, we study the case where $\alpha=2$, which gives:
\be
\phi=\sqrt{\frac{R-2k\pm\sqrt{R(1-2k)}}{2}}\ .
\label{1.19}
\ee  
By inserting this expression into (\ref{1.13}) and (\ref{1.15}), the function $f(R)$ is obtained:
\be
f(R)=\left[R-6(k(k+1)+5/2) \right]\frac{R-2k\pm\sqrt{R(1-2k)}}{2}+const\ .
\label{1.20}
\ee
Thus, with this expression for the function $f(R)$, the current cosmic acceleration is reproduced with the solution (\ref{1.16}). In general, as it is seen in the following example, it is very difficult to reconstruct the function $f(R)$ for the whole expansion history, and even more difficult for the kind of models that unify inflation and cosmic acceleration, in this cases it is convenient to study the asymptotic behaviour of the model, and then by resolving the equations, the expression for $f(R)$ is obtained .\\ \\

As a second example, one could try to reconstruct the whole Universe  history, from inflation to  cosmic acceleration by the f(R)-gravity as is made in \cite{F(R)toScalar} -\cite{F(R)UnfInfCosAcce4}. In this case, we proceed in the inverse way than above, by suggesting a function $g(\phi)$, and  trying to reconstruct the expression $f(R)$ by calculating $P(\phi)$ and $Q(\phi)$ by means of equations (\ref{1.8}). We study an example suggested in \cite{MyW1}, where:
\be
g(\phi)=\frac{H_1}{\phi^2}+\frac{H_0}{t_s-\phi}\ .
\label{1.21}
\ee
For this function the Hubble parameter takes the form $H(t)=\frac{H_1}{t^2}+\frac{H_0}{t_s-t}$. To reconstruct the form of $f(R)$ we have to resolve the equation (\ref{1.10}). For simplicity we study the assimptotically behaviour of $H(t)$ in such a way that allow us to resolve easily the equation (\ref{1.10}) for $P(\phi)$. Then, for small $t$ ($t<<t_s$), the Hubble and acceleration parameters read:
\be
H(t)\sim\frac{H_1}{t^2}\, \quad \frac{\ddot{a}}{a}\sim \frac{H_1}{t^2}\left(\frac{H_1}{t^2}-\frac{2H_1}{t}+\frac{2H_0}{t_s-t} \right)\ .
\label{1.22}
\ee
As it is observed, for t close to zero, $\frac{\ddot{a}}{a}>0$, so the Universe is in an accelerated epoch during some time, which may be interpreted as the inflation epoch, and for $t>1/2$ ($t<<t_s$), the Universe enters in a decelerated phase, interpreted as the radiation/matter dominated epoch. The equation (\ref{1.10}) for $P(\phi)$, neglecting the matter component, is given by:
\be
\frac{d^2P(\phi)}{d\phi^2}-\frac{H_1}{×\phi^2}\frac{dP(\phi)}{d\phi}-\frac{4H_1}{\phi^3}P(\phi)=0\ .  
\label{1.23}
\ee
The solution of (\ref{1.23}) is:
\be
P(\phi)=k\phi^{-4}+\frac{20k}{×H_1}\phi^{-3}+\frac{120k}{×H^2_1}\phi^{-2}+\frac{240k}{×H^3_1}\phi^{-1}+\frac{120k}{×H_1^4}\ , 
\label{1.24}
\ee
where $k$ is an integration constant. The function $Q(\phi)$ given by (\ref{1.11}), takes the form:
\be
Q(\phi)=-6\frac{H_1}{×\phi^2}\left[kH_1\phi^{-6}+6k\phi^{-4}-\frac{120k}{×H^{3}_1}\phi^{-2} \right]\ .
\label{1.25}
\ee
By using  the expression  for the Ricci scalar, the relation $\phi(R)\sim(12H^2_1/R)^{1/4}$ is found,  then, the function $f(R)$ for small values of $\phi$ is approximately:
\be
f(R)\sim\frac{k}{24H_1}R^2\ .
\label{1.26}
\ee
Hence, by the expression (\ref{1.26}) the early cosmological behaviour of the Universe (\ref{1.22}), where a first accelerated epoch (inflation) occurs and after it,   a decelerated phase comes (radiation/matter dominated epochs),  is reproduced. Let us now investigate the large values for $t$ ($t$ close to the Rip time $t_s$). In this case the Hubble and the acceleration parameters for the solution (\ref{1.21}) take the form:
\be
g(\phi)\sim\frac{H_0}{×t_s-\phi} \quad \rightarrow H(t)\sim\frac{H_0}{×t_s-t}\ \quad \mbox{and} \quad \frac{\ddot{a}}{×a}\sim \frac{H_0(H_0+1)}{×(t_s-t)^2}\ .
\label{1.27}
\ee
As it is observed, for large $t$ the solution (\ref{1.21}) gives an accelerated epoch which enters in a phantom phase ($\dot{H}>0$) and ends in a Big Rip singularity at $t=t_s$. In this case the equation for $P(\phi)$ reads:
\be
\frac{d^2P(\phi)}{d\phi^2}-\frac{H_0}{×t_s-\phi}\frac{dP(\phi)}{d\phi}+\frac{2H_0}{(t_s-\phi)^2}P(\phi)=0\ .
\label{1.28}
\ee
The possible solutions of the equation (\ref{1.28}) depend on the value of the constant $H_0$, as it follows:
\begin{enumerate}

 \item If $H_0>5+2\sqrt{6}$ or $H_0<5-2\sqrt{6}$, then the following solution for $P(\phi)$ is found: 
\bea
P(\phi)=A(t_s-\phi)^{\alpha_+}+B(t_s-\phi)^{\alpha_-}, \nn 
\mbox{where} \quad \alpha_{\pm}=\frac{H_0+1\pm\sqrt{H_0(H_0+10)+1}}{×2}\ ,
\label{1.29}
\eea
Then, through the expression (\ref{1.11}), the function $Q(\phi)$ is calculated. In this case, for $t$ close to $t_s$, the Ricci scalar takes the form $R=\frac{6H_0(2H_0+1)}{×(t_s-t)^2}$, and hence it takes large values, diverging at the Rip time $t=t_s$. The function $f(R)$, for a large $R$, takes the form:
\be
f(R)\sim R^{1-\alpha_-/2}\ .
\label{1.30}
\ee

\item If  $5-2\sqrt{6}<H_0<5+2\sqrt{6}$, the solution of (\ref{1.28}) is given by:
\[
 P(\phi)=(t_s-\phi)^{-(h_0+1)/2}\left[A\cos\left((t_s-\phi)\ln\frac{-H^2_0+10H_0-1}{2×} \right) \right.
\]

\be
\left.
+B\sin\left((t_s-\phi)\ln\frac{-H^2_0+10H_0-1}{2×} \right)  \right]\ .
\label{1.31}
\ee
Then, for this choice of the constant $H_0$, and by means of the equation (\ref{1.7}) the form of the function $f(R)$ is found:
\[
 f(R)\sim R^{(H_0+1)/4}\left[A\cos\left(R^{-1/2}\ln\frac{-H^2_0+10H_0-1}{2×} \right) \right.
\]
\be
 \left. +B\sin\left(R^{-1/2}\ln\frac{-H^2_0+10H_0-1}{2×} \right)  \right]
\ .
\label{1.32}
\ee
\end{enumerate}
Hence, the expressions (\ref{1.30}) and (\ref{1.32}) for $f(R)$ reproduce the behaviour of the Hubble parameter for large $t$ given in (\ref{1.27}), where a phantom accelerated epoch ocurrs, and the Universe ends in a Big Rip singularity for $t=t_s$. As  is shown, this model is reproduced by (\ref{1.26}) for small $t$ when the curvature $R$ is large, and by (\ref{1.30}) or (\ref{1.32}) when $t$ is large. For a proper choice of the power $\alpha$, the solution for large $t$ is given by (\ref{1.30}), which in combination with the solution (\ref{1.26}) for small $t$, it looks like standard gravity $f(R)\sim R$ for intermediate $t$. On the other hand, for negavite powers in (\ref{1.30}) and in combination with (\ref{1.26}), this takes a similar form than the model suggested in \cite{f(R)4}, $f(R)\sim R+R^2+1/R$, which is known that passes qualitatively most of the solar system tests. As this is an approximated form, it is reasonable to think that this model follows from some non-linear gravity of the sort Ref.\cite{F(R)UnfInfCosAcce3}, which  may behave as $R^2$ for large $R$. The stability of this kind of models (for a detailed discussion see \cite{StabilityF(R)}), whose solutions are given by (\ref{1.26}) and (\ref{1.32}), is  studied in Ref. \cite{F(R)toScalar}, where the transition between epochs is well done, and then, the viable cosmological evolution may be reproduced by these models.  The quantitative study of the transition between different cosmological epochs is beyond of the purposes of this work.       \\ \\
Suming up it has been shown that any cosmology may be reproduced by $f(R)$-gravity by using an auxiliary scalar field and resolving the equations (\ref{1.10}) and (\ref{1.11}) to reconstruct such function of the Ricci scalar. To fix the free parameters in the theory, it would be convenient to contrast the model with the observational data as the supernova data by means of the evolution of the scale parameter which is obtained in the models showed above.

\section{F(R)-gravity and dark fluids}

In this section  the mathematical equivalence between f(R) theories, that could reproduce a given cosmology as it was seen above, and the standard cosmology with a dark fluid included whose EoS has inhomogeneous terms that depend on the Hubble parameter and its derivatives, is investigated. Let us start with the modified Friedmann equations (\ref{1.3}) written in the following form:
\bea
3H^2=\frac{1}{f'(R)×}\left(\frac{1}{2×}f(R)+\nabla_0\nabla^0f'(R)- \Box f'(R)\right) -3\dot{H}\ , \nn
-3H^2-2\dot{H}=\frac{1}{f'(R)×}\left(\Box f'(R)-\frac{1}{2×}f(R)-\dot{H} \right)\ ,
\label{2.1}
\eea
where we have neglected the contributions of any other kind of matter. If we compare Eqs. (\ref{2.1}) with the standard Friedmann equations ($3H^2=\kappa^2\rho$ and $-3H^2-2\dot{H}=\kappa^2p$), we may identify both right sides of Eqs. (\ref{2.1}) with the energy and pressure densities of a perfect fluid, in such a way that they are given by:
\bea
\rho=\frac{1}{\kappa^2×}\left[ \frac{1}{f'(R)×}\left(\frac{1}{2×}f(R)+\nabla_0\nabla^0f'(R)- \Box f'(R)\right) -3\dot{H}\right]  \nn
p=\frac{1}{\kappa^2×}\left[\frac{1}{f'(R)×}\left(\Box f'(R)-\frac{1}{2×}f(R)\right)-\dot{H}\right]\ . 
\label{2.2}
\eea
Then, Eqs. (\ref{2.1}) take the form of the usual Friedmann equations, where the parameter of the EoS for this dark fluid is defined by:
\be
w=\frac{p}{\rho×}=\frac{\frac{1}{f'(R)×}\left(\Box f'(R)-\frac{1}{2×}f(R)\right)-\dot{H}}{\frac{1}{f'(R)×}\left(\frac{1}{2×}f(R)+\nabla_0\nabla^0f'(R)- \Box f'(R)\right) -3\dot{H}}\ .
\label{2.3}
\ee
And the corresponding EoS  may be written as follows:
\be
p=-\rho-4\dot{H}-\frac{1}{f'(R)}\nabla_0\nabla^0f'(R)\ .
\label{2.4}
\ee
The Ricci scalar is a function given by $R=6(2H^2+\dot{H})$, then $f(R)$ is a function on the Hubble parameter $H$ and its derivative $\dot{H}$. The inhomogeneus EoS for this dark fluid (\ref{2.4}) takes the form of the kind of dark fluids studying in several works (see \cite{InhEoS1}-\cite{MyW2}), and particularly the form of the EoS for dark fluids investigated in \cite{MyW2}, which is written as follows:
\[
p=-\rho+g(H,\dot{H}, \ddot{H}...), \quad \mbox{where}     
\]
\be
g(H,\dot{H}, \ddot{H}...)=-4\dot{H}+\nabla_0\nabla_0(\ln f'(R))+(\nabla_0\ln f'(R))(\nabla_0\ln f'(R))\ .
\label{2.5}
\ee
Then, as  constructed in \cite{MyW2}, by combining the Friedmann equations, it yields the following differential equation:
\be
\dot{H}+\frac{\kappa^2}{2}g(H,\dot{H}, \ddot{H}...)=0\ .
\label{2.6}
\ee
Hence, for a given cosmological model, the function $g$ given in (\ref{2.5}) may be seen as a function of cosmic time $t$, and then by the time-dependence of the Ricci scalar, the function $g$ is rewritten in terms of $R$. Finnally, the function $f(R)$ is recovered by the expression (\ref{2.5}). In this sense, Eq. (\ref{2.6}) combining with the expression (\ref{2.5}) results in:
\be
\frac{dx(t)}{dt}+x(t)^2=\dot{H}(t)\ ,
\label{2.7}
\ee
where $x(t)=\frac{d(lnf'(R(t))}{dt}$. Eq. (\ref{2.7}) is a type of Riccati equation, that may be solved by a given Hubble parameter. Hence,  $f(R)$-gravity is constructed from standard cosmology where a perfect fluid with an inhomogeneus EoS is included. To show this, let us consider the following example:
\be
H(t)=H_0t+\frac{H_1}{t}\ , \quad H_1,H_0>0\ . 
\label{2.8}
\ee
This model reproduces a Universe that passes through two epochs, a first decelerated phase and a second one accelerated that is identified with the current epoch. To resolve (\ref{2.7}) for this example, and for simplicity, we study the asymptotic behaviour of the model. Then, for small $t$, the Hubble parameter takes the form $H(t)\sim\frac{H_1}{t}$, and the form of the function $f(R)$ is found by solving Eq. (\ref{2.7}), to be:
\be
f(R)\sim\frac{1}{1-k/2}R^{1-k/2}+\lambda\ ,
\label{2.9}    
\ee
where $k(1-k)=H_1$ and $\lambda$ is an integration constant. Then, by the function (\ref{2.9}) the model (\ref{2.8}) for small $t$ is reproduced. The analog dark fluid that reproduces this behaviour, may be found by inserting the function (\ref{2.9}) in the EoS for the fluid given by (\ref{2.5}), it yields: 
\be
p\sim-\rho+\frac{1}{\kappa^2}\left[-4\dot{H}+\frac{1}{R}\left(k/2(1+k/2)\frac{\dot{R}}{R}-\frac{k}{2}\ddot{R}\right)\right]\ ,
\label{2.10}
\ee
where $R=6(2H^2+\dot{H})$. Then, a perfect fluid with an inhomogeneous EoS given by (\ref{2.10}) reproduces the asymptotic behaviour for small $t$ in an analog description to $f(R)$. Let us now explore the form of $f(R)$ function and the EoS of the dark fluid for large $t$, in this case $H(t)\sim H_0 t$, and by resolving the equation (\ref{2.7}), the expression for $f(R)$ is given by:
\be
f(R)\sim\exp{\sqrt{H_0R}}\left(2\sqrt{\frac{R}{H_0}}-\frac{2}{H_0}\right)+\lambda\ ,
\label{2.11}
\ee
where $\lambda$ is an integration constant. The alternative description in terms of a dark fluid is shown trough its  EoS, which is calculated as in the case above, by the equation (\ref{2.5}):   
\be
p\sim-\rho+\frac{1}{\kappa^2}\left[-4\dot{H}+\frac{\dot{R}^2}{H_0R}(1+\sqrt{\frac{H_0}{4R}})-\frac{\ddot{R}}{\sqrt{H_0R}}\right]\ ,
\label{2.12}
\ee
which gives an EoS dependent on the Hubble parameter and higher derivatives contained in the Ricci scalar $R=6(2H^2+\dot{H})$. Hence, the example considered (\ref{2.8}) is reproduced in $f(R)$-gravity, where its asymptotical behaviour is given  by the functions calculated above ((\ref{2.9}) and (\ref{2.11})). And  the same result may be reproduced by a dark fluid whose EoS is given by the functions (\ref{2.10}) and (\ref{2.12}). Thus, as it is showed,  f(R)-gravity may be written as dark fluids with particular dependence on the Hubble parameter and its derivatives through its EoS (\ref{2.5}), so the same model may be interpretated in several ways.
 
\section{Discussions}
f(R)-gravity theories may provide an alternative description of the current accelerated epoch of our Universe and even on the whole expansion history. As it is pointed in several works (see \cite{f(R)viable1}-\cite{f(R)viable10} ), one may construct this kind of theories in accordance with the local test of gravity and with the observational data, which provide that, at the current epoch, the effective parameter of the EoS is close to -1. The next step should be to compare the different cosmological tests,  as the supernovae luminosity distance or the positions of the CMB peaks, with the F(R) models,  but this is beyond of the purpose of this paper. In other hand, we have shown two different ways of reconstruct the f(R)-gravity in the context of cosmology, in the first one, an auxiliary scalar field is used, and in the second one, the mathematical equivalence between f(R)-gravity and dark fluids with inhomogeneus EoS  shows  that while the expansion history of the Universe may be interpreted as a perfect fluid whose EoS has dependence on the cosmological evolution, this effect may be caused by the modification of the classical theory of gravity. However, there is not any complementary probe to distinguish between both descriptions of the evolution of the Universe, and thus, such kind of modified gravity is completly allowed. Hence, f(R)-gravity is an acceptable solution to the cosmological problem, that may provide new interesting constraints to look for.        

\begin{acknowledgements}

 I thank  Emilio Elizalde and Sergei Odintsov for suggesting this problem, and for giving the ideas and fundamental information to carry out this task. This work was supported  by MEC (Spain), project FIS2006-02842, and in part by project PIE2007-50/023.

\end{acknowledgements}


\begin{thebibliography}{}
\bibitem{f(R)review1}S. Nojiri and S. D. Odintsov, Int. J. Geom. Meth. Mod. Phys. \textbf{4} 115 (2007) [arxiv:hep-th/0601213]
\bibitem{f(R)review2}Thomas P. Sotiriou, Valerio Faraoni, [arXiv:0805.1726 [gr-qc]]

\bibitem{f(R)1}S. Capozziello, Int. J. Mod. Phys. D \text{11}, 483 (2002)
\bibitem{f(R)3}S. M. Carroll, V. Duvvuri, M. Trodden and S. Turner, Phys. Rev. D 70 (2004) 043528.
\bibitem{f(R)4}S. Nojiri and S. D. Odintsov, Phys. Rev. D \textbf{68}, 123512 (2003) [arXiv:hep-th/0307288]. 
\bibitem{f(R)6}M. Ruggiero and L. Iorio, arXiv:gr-qc/0607093; A. Cruz-
\bibitem{f(R)7}Dombriz and A. Dobado, arXiv:gr-qc/0607118;  
\bibitem{f(R)9}A. Brookfield, C. van de Bruck and L. Hall, arXiv:hep-th/0608015; 
\bibitem{f(R)17}G. Olmo, arXiv:gr-qc/0612047; 
\bibitem{f(R)18}F. Briscese, E. Elizalde, S. Nojiri and S. D. Odintsov, Phys. Lett.B 646, 105 (2007) [arXiv:hep-th/0612220]; 
\bibitem{f(R)19}B. Li and J. Barrow, arXiv:gr-qc/0701111; 
\bibitem{f(R)21}V. Faraoni, arXiv:gr-qc/0703044; arXiv:0706.1223 
\bibitem{f(R)23}O. Bertolami,C. Boehmer, T. Harko and F. Lobo, arXiv:0704.1733; 
\bibitem{f(R)24}S. Carloni, A. Troisi and P. Dunsby, arXiv:0706.0452; 0707.0106;
\bibitem{f(R)26}S. Capozziello and M. Francaviglia, arXiv:0706.1146.
\bibitem{f(R)27}S. Nojiri and S. Odintsov, Gen. Rel. Grav. 36, 1765 (2004) [arXiv:hep-th/0308176]; Phys. Lett. B\textbf{576} 5 (2004)[arXiv:hep-th/0307071]; 
\bibitem{f(R)29}M. Abdalla, S. Nojiri and S. D. Odintsov, Class. Quant. Grav. 22, L35 (2005) [arXiv:hep-th/0409177]; 
\bibitem{f(R)30}G. Cognola, E. Elizalde, S. Nojiri, S. D. Odintsov and S. Zerbini, JCAP 0502, 010 (2005) [arXiv:hep-th/0501096]; Phys. Rev. D73, 08400(2006), [arXiv:hep-th/0601008]; 
\bibitem{f(R)32}S. Capozziello, V. Cardone and A. Troisi, arXiv:astro-ph/0501426; 
\bibitem{f(R)34}G. Allemandi, M. Francaviglia, M. Ruggiero and A. Tartaglia, arXiv:gr-qc/0506123; 
\bibitem{f(R)35}T. Multamaki and I. Vilja, arXiv:astro-ph/0612775; 
\bibitem{f(R)36}J. A. R. Cembranos, Phys. Rev. D 73, 064029 (2006) [arXiv:gr-qc/0507039]; 
\bibitem{f(R)38}T. Clifton and J. Barrow, arXiv:gr-qc/0509059; 
\bibitem{f(R)39}O. Mena, J. Santiago and J. Weller, arXiv:astro-ph/0510453; 
\bibitem{f(R)41}I. Brevik, arXiv:gr-qc/0601100; 
\bibitem{f(R)42}R.Woodard, arXiv:astro-ph/0601672; 
\bibitem{f(R)45}G. Cognola, M. Castaldi and S. Zerbini, arXiv:gr-qc/0701138; 
\bibitem{f(R)46a}S. Nojiri,S. D. Odintsov and P. Tretyakov, arXiv:0704.2520[hep-th]; 
\bibitem{f(R)47}M. Movahed, S. Baghram and S. Rahvar, arXiv:0705.0889[astroph];
\bibitem{f(R)48}L. Amendola and S. Tsujikawa, arXiv:0705.0396[astro-ph]; 
\bibitem{f(R)51}C. Boehmer, L. Hollenstein and F. Lobo, arXiv:0706.1663.
\bibitem{f(R)53}S. Capozziello, S. Nojiri, S. D. Odintsov and A. Troisi, Phys. Lett. B 639, 135 (2006).
\bibitem{f(R)54}S. Fay, S. Nesseris and L. Perivolaropoulos, arXiv:gr-qc/0703006; 

\bibitem{F(R)toScalar} S. Nojiri, S. D. Odintsov, Phys. Rev. D \textbf{74} (2006) 086005 [arXiv:hep-th/0608008]; hep-th 0611071
\bibitem{F(R)UnfInfCosAcce2andsolartest}S. Nojiri, S. D. Odintsov, Phys. Lett. B \textbf{657} 238-245 (2007) [arXiv:hep-th/0707.1941]
\bibitem{F(R)UnfInfCosAcce3}S. Nojiri, S. D. Odintsov, Phys. Rev. D \textbf{77} 026007 (2008) [arXiv:0710.1738[hep-th]]
\bibitem{F(R)UnfInfCosAcce4}G.Cognola, E. Elizalde, S. Nojiri, S. D. Odintsov, L. Sebatiani and S. Zerbini, Phys. Rev. D \textbf{77} 046009 (2008) [arXiv:0712.4017[hep-th]]
\bibitem{F(R)UnfInfCosAcceAndSingularity}S. Nojiri, S. D. Odintsov, arxiv:0804.3519 [hep-th]
\bibitem{f(R)viable1} W. Hu and I. Sawicki, arXiv:0705.1158[astro-ph].
\bibitem{f(R)viable2} S. A. Appleby and R. A. Battye, arXiv:0705.3199[astro-ph];
\bibitem{f(R)viable3}L. Pogosian and A. Silvestri, arXiv:0709.0296[astro-ph];
\bibitem{f(R)viable4}S. Tsujikawa, arXiv:0709.1391[astro-ph];
\bibitem{f(R)viable5}A. Starobinskii, JETP Lett.86 157 (2007);
\bibitem{f(R)viable6}S. Capozziello and S. Tsujikawa, arXiv:0712.2268[gr-qc].
\bibitem{f(R)viable7}S. Nojiri and S. D. Odintsov, arXiv:0706.1378[hep-th]; 
\bibitem{f(R)viable9}G. Cognola, E. Elizalde, S. Nojiri, S. D. Odintsov, L. Sebastiani and S. Zerbini, Phys. Rev. D \textbf{77} 046009 (2008) [arxiv:0712.4017 [hep-th]]
\bibitem{f(R)viable10}S. Nojiri and S. D. Odintsov, arXiv:0801.4843[astro-ph]
\bibitem{Quintessence1} M. Sahl\'{e}n, A. R. Liddle and D. Parkinson, Phys. Rev. D \textbf{72} 083511 (2005) [arxiv: astro-ph/050669]
\bibitem{Quintessence2} Chao Li, Daniel E. Holz, A. Cooray, Phys. Rev. D \textbf{75}  103503 (2007) [arxiv: atro-ph/0611093]
\bibitem{InhEoS1} S. Nojiri and S.D.Odintsov, Phys. Rev. D \textbf{72}, 103522 (2005) [arxiv:hep-th/0505215]; 

\bibitem{InhEoS2}S. Capozziello, V. Cardone, E. Elizalde, S. Nojiri and S.D. Odintsov, Phys. Rev. D \textbf{73}, 043512 (2006), [astro-ph/0508350]

\bibitem{InhoEoSandCoupling} S. Nojiri and S. D. Odintsov, Phys. Lett. B \textbf{639}, 144 (2006), [arxiv:hep-th/0606025]

\bibitem{InhEoSandOscillating1} I. Brevik, O.G. Gorbunova and A. V. Timoshkin, Eur.Phys.J.C51:179-183,(2007) [arxiv:gr-qc/0702089];
\bibitem{InhEoSandOscillating2}
I.Brevik, E. Elizalde, O. Gorbunova. and A. V. Timoshkin Eur. Phys. J. C \textbf{52} 223 (2007) [arxiv:gr-qc/0706.2072]
\bibitem{MyW2}Diego S\'{a}ez-G\'{o}mez, [arxiv:0804.4586 [hep-th]]
\bibitem{MyW1}E. Elizalde, S. Nojiri, S. D. Odintsov, D. S\'{a}ez-G\'{o}mez and V. Faraoni, Phys. Rev. D \textbf{77} 106005 (2008) [arxiv:hep-th/0803.1311]
\bibitem{StabilityF(R)} Luca Amendola, R. Gannouji, D. Polarski and S. Tsujikawa, Phys. Rev. D \textbf{75} 083504 (2007) [arxiv: gr-qc/0612180]
\end{thebibliography}
\end{document}